\documentclass[12pt]{iopart}

\usepackage{graphicx}
\usepackage{url}

\usepackage{siunitx}
\DeclareSIUnit\gauss{G}

\usepackage{soul}
\usepackage{color}
\definecolor{bleuclair}{rgb}{0.7, 0.7, 1.0}
\definecolor{rosepale}{rgb}{1.0, 0.7, 1.0}

\begin{document}

\title{Low-Cost Experiments with Everyday Objects for Homework Assignments}

\author{F. Bouquet$^1$,  
       C. Dauphin$^2$,
       F. Bernard$^3$,
       and J. Bobroff$^1$}

   \address{$^1$ Laboratoire de Physique des Solides, CNRS, University Paris-Sud, Universit\'e Paris-Saclay,
91405 Orsay cedex, France}
   \address{$^2$ Institut Villebon-\textit{Georges Charpak}, 91400 Orsay, France}
   \address{$^3$ Institut d'Optique, Graduate School, 91120 Palaiseau, France}

\begin{abstract}
We describe four classical undergraduate physics experiments that were done with everyday objects and low-cost sensors: mechanical oscillations, transmittance of light through a slab of matter, beam deformation under load, and thermal relaxation due to heat loss. We used these experiments to train students for experimental homework projects but they could be used and expanded in a variety of contexts: lecture demonstrations, low cost students' labs, science projects, distance learning courses...
\end{abstract}

\maketitle

\section{Introduction}

One often neglected benefit of low-cost experiments is that they can be taken outside of the classroom with little financial risk. We gave second-year university science students experimental homework assignments: students had to carry a physical study at home, using Arduino~\cite{refArduino} and associated sensors. The objective was to develop their scientific practices of experimental physics.  Far from a cookbook procedure, they had to decide how to perform the study, what to measure, and how to interpret the results. Using everyday objects instead of specific lab materials shows students that physics is everywhere, not just in  labs.
Each experiment costs less than 40~euros (half of it for the cost of the board).  The low-cost microcontroller Arduino was not developed as a physicist's  tool; nevertheless, it has been recently used in various context of experimental physics activities~\cite{refKubinova,refGaleriu,refAtkin,refPetry,refHenaff,refBouquet}: its ease-of-use, low cost, and large user community make it a good tool to develop new experimental set-ups~\cite{refOpenTP}, even though its specifications are somewhat limited compared to other low-cost microcontrollers~\cite{refMbed}, which could be used instead if more precise measurements are needed.

The students were already familiar with the Arduino board but only as a tool for tinkering projects, having been subject to a two-day workshop using Arduino challenge sheets~\cite{refOpenTP} in their first year.
Prior to this experimental homework assignment, we trained the students in using sensors with the Arduino board and in conducting an experimental study (building the setup, gathering and analyzing data). 

In this article, we describe the four low-cost Arduino-controlled experiments that we used for this training, each experiment corresponding to a 3-hour in-class session. These experiments can easily be reproduced or modified, and cover a wide range of physics.

\section{Mechanical oscillations}

\textit{Sensor:} accelerometer

\noindent \textit{Experiment:} damped harmonic motion of a mass suspended to an elastic

This experiment is used to introduce the students to Arduino as a scientific measuring device.
The experimental setup is simple and easy: the accelerometer is fixed to a massive object suspended to a long elastic band, or to a spring. It is easy to create an elastic band of any chosen length by chaining common office rubber bands together.

The object is then pulled down a bit, and let go. The readings of the accelerometer can be analysed by comparison to the harmonic oscillator model (with or without friction). Special care should be taken so that the wires connecting the accelerometer to the board are flexible enough and do not damp the oscillations, and the setup should minimize spurious pendulum-like oscillations. For example, using a bottle of water as a mass is not advised, as slushing gives a signal that is easily picked by the sensor. For simplicity, the accelerometer should also be positioned so that one of its measurement axis lies parallel to the oscillation direction, which is a point students often miss.

Many accelerometers can be read by an Arduino board~\cite{refAccelerometer}, and generally cost about 10~euros. They use microelectromechanical inertial sensors, which by construction measure the acceleration of the sensor minus the acceleration of gravity $\vec{g}$. Thus calibrating the accelerometer is easily done when the system is at rest, since the reading of the sensor should be  \SI{+-9.8}{\meter\per\second\squared} when the direction of the sensor is vertical (the sign depends on the direction), and  \SI{0}{\meter\per\second\squared} when the direction of the sensor is horizontal.

\begin{figure}%[th]
\begin{center}
\includegraphics[width=0.45\textwidth]{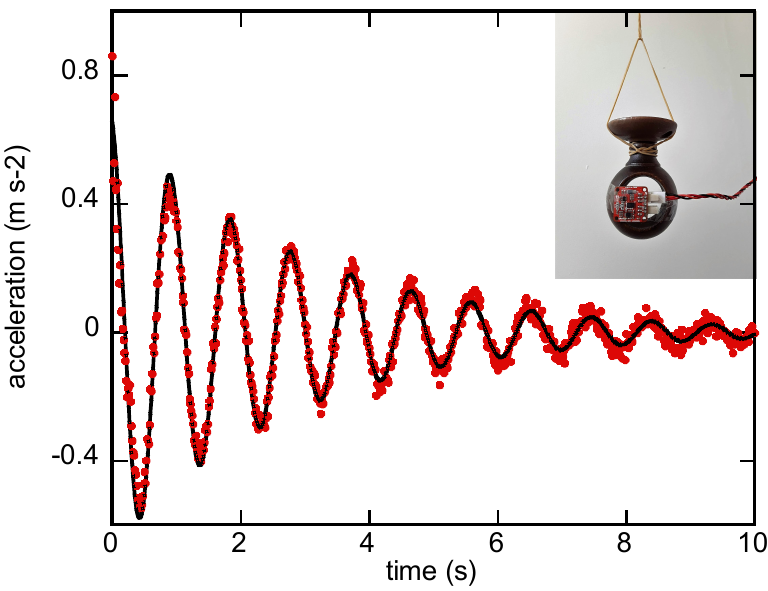}
\end{center}
\caption{Oscillations of a \SI{160}{\gram} curtain-rod tip suspended to 15 office rubber bands chained together in a one-meter long elastic system. The data points represent the acceleration of the object, $g$ has been subtracted from the readings of the accelerometer. The solid line is a fit of the data, with the initial deformation fixed at \SI{1.5}{\centi\meter}, and with $\tau$ and $\omega$ as free parameters. The best fit is given by $\tau = \SI{2.8}{\second}$ and $\omega =  \SI{6.7}{\radian\per\second}$. } \label{figure_accelerometre}
\end{figure}

Figure~\ref{figure_accelerometre} shows typical results for such setup: vertical oscillations that are rapidly damped. During the first session with students, the damping is not analysed: only the angular frequency is determined by direct measurement and neglecting the effect of the damping, but the complete study is one of the home assignments that students can choose. 
The data can be analyzed using the equation of the movement of an underdamped oscillator:
     \begin{equation}
     z(t)= A \exp^{-\frac{t}{\tau}}\cos (\omega t + \Phi_0) \quad,
     \label{eq_accelerometre}
     \end{equation}
where $z$ is the vertical position of the object, $A$ is the  amplitude of the oscillation, $\tau$ is the damping time, $\omega$ the angular frequency, and $\Phi_0$ is the initial phase of the oscillation. The angular frequency is related to the undamped angular frequency $ \omega_0 = \sqrt{k/m} $ by the relation $ \omega = \omega_0 \sqrt{1 - 1/\tau^2\omega_0^2} $ (with $k$ the elastic constant and $m$ the mass of the object). 

By pulling the system out of equilibrium to a given position $A$, and letting it go, the measured acceleration can be fitted using the second derivative of equation~\ref{eq_accelerometre} with only two free parameters, $\tau$ and $\omega$ (choosing $t=0$ when the object is let go defines $\Phi_0 = 0$ if the damping is small, but a small $\Phi_0$ can be added as a third free parameter if not, or to accommodate an uncertainty on the $t=0$ moment). Results of the fit are presented in Fig~\ref{figure_accelerometre}.

Further experiments can investigate the role of the total length of the elastic, or the effect of putting several elastic bands in parallel.  

\section{Transmission of light}
\textit{Sensor:} light sensor

\noindent \textit{Experiment:} measuring the quantity of light going through several layers of transparent sheets.

The goal of this session is to let the students work on scientific graphs, and on how to present data. The experiment consists in measuring the quantity of light going through several layers of transparent sheets and studying how this quantity depends on the number of sheets. Many light sensors can be used, but to achieve a better precision  it is best to avoid uncalibrated analog sensor of which linearity may not be guaranteed~\cite{refAnalogLightSensor}, and to prefer digital calibrated sensor, or the light sensor of a smartphone with the appropriate app. 

\begin{figure}%[th]
\begin{center}
\includegraphics[width=0.45\textwidth]{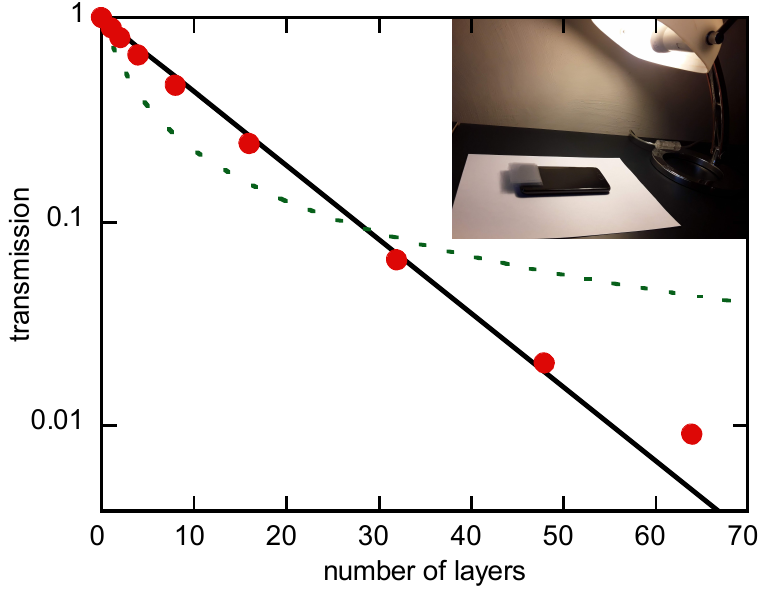}
\end{center}
\caption{Transmission coefficient of a stack of transparent sheets. For this experiment, the transparent sheets where sheets proposed as cover in the local reprography center, and a smartphone sensor was used with the phyphox application\cite{refPhyphox}. The light source was a simple desk light (see inset). The dashed line represents the multiple reflections/transmissions (in a multilayer system) model. The straight line is the exponential model. } \label{figure_lumiere}
\end{figure}

The experiment is straightforward, but should be done with care, with a light source that does not vary with time and does not saturate the sensor. Ambient light or desktop light can be used, as long as the total thickness of the sheets does not become too large, in which case some leakage of light can appear through the sides. Any transparent sheets can be used: plastic book covers, wrappings, \ldots (the setup is shown in the inset of figure~\ref{figure_lumiere}).

A typical result is given in figure~\ref{figure_lumiere}, using a smartphone and phyphox app~\cite{refPhyphox}  to measure the transmitted light.

Two different models can be candidates for describing this experiment. A first model considers that the light rays are always normal to the sheets, and that the physics at play in the stack of sheets is multiple reflections and transmissions in a multilayer system (neglecting diffusion and absorption within the sheets). By defining $T_1$ and $R_1$ the coefficient of transmission and reflection of a single sheet (with $R_1 + T_1 = 1$), and $N$ the number of sheets, mathematical induction shows that the total transmission of light $T_N$ can be written as:
    \begin{equation}
    T_N = \frac{1-R_1}{1+(N-1)R_1} \quad.
     \end{equation}

A second model considers scattering or absorption of light within the sheets, and the resulting transmission coefficient will follow the Lambert law:
    \begin{equation}
    T_N = \exp(-\alpha N) \quad,
     \end{equation}
where $\alpha$ characterizes the transmission of a single sheet, $T_1=\exp(-\alpha)$.

Both models predict a different dependence of the total transmittance of the plastic-sheet stack in function of the number of sheets. Figure~\ref{figure_lumiere} shows the best adjustment  of these models to the data. Clearly the Lambert's law gives a better agreement, with $\alpha = \num{8.35e-2}$ per sheet: even though the plastic is very clear and flat, the simple model of normal rays being only reflected or transmitted at each layer does not encompass the physics at play.

%scattering and absorption seem to dominate the behavior of the stack. 
In these data, the 64-layer point deviates from the model, either because the limit of sensitivity of the sensor is reached, or because the stack becomes too thick and light can reach the sensor through the sides.

Further studies can investigate the role of the wavelength of the illuminating source and the effect of using  colored sheets instead of transparent ones. More elaborate setups can be imagined~\cite{refSpectro1,refSpectro2}.

Light sensors can also be used to study the amount of light received in function of the distance between the source and the sensor, or to track a movement through the shadows of the object of interest; both were proposed as home assignments.

\section{Deformation of a ruler}
\textit{Sensor:} linear Hall sensor

\noindent \textit{Experiment:} deformation of a beam with central load

\begin{figure}%[th]
\begin{center}
\includegraphics[width=0.45\textwidth]{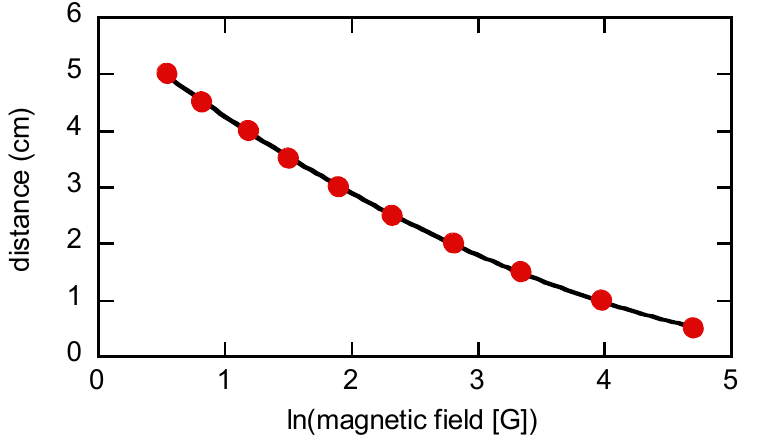}
\end{center}
\caption{Calibration of the magnetic field of a NdFeB magnet with distance. The calibration is more precise when performed in function of the logarithm of the magnetic field (in gauss here), since the value of the field varies over two decades with distance. In this experiment, a second order polynomial fit is sufficient to describe the calibration points.} \label{figure_calibration}
\end{figure}

This experiment is used to introduced the students to the notion of fitting data and using a calibration. To do so, the Hall sensor is not used to study a magnetostatic phenomenon, but to measure a distance. Prior to the main experiment, the field produced by a magnet is carefully measured in function of the distance between the sensor and the magnet, and a fitting procedure is performed to convert the sensor reading into a distance (see figure~\ref{figure_calibration}). As always with Hall sensor, special care should be taken for the orientation of the sensor vis-\`a-vis the magnet. This way of measuring distance can be quite sensitive for small distance variations when the magnet is close to the sensor; the range will depend on the choice of sensor and magnet (the derivative of the calibration equation gives the sensitivity of the setup: in our case the sensitivity was  \SI{70}{\gauss\per\milli\meter} at \SI{1}{\centi\meter},  \SI{16}{\gauss\per\milli\meter} at \SI{2}{\centi\meter}, and \SI{6}{\gauss\per\milli\meter} at \SI{3}{\centi\meter}).

\begin{figure}%[th]
\begin{center}
\includegraphics[width=0.45\textwidth]{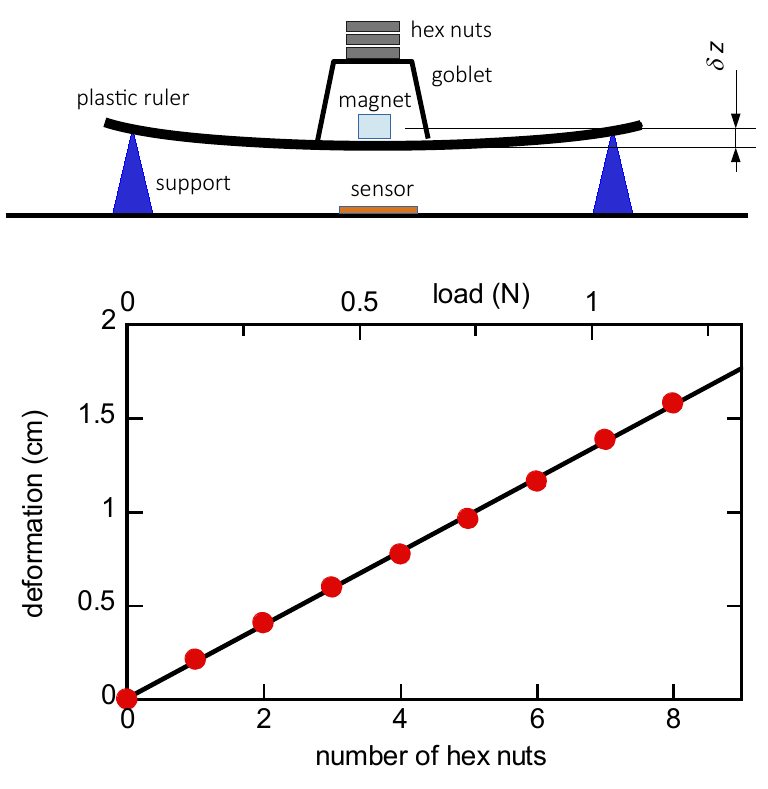}
\end{center}
\caption{Top figure: experimental setup. A plastic  ruler is placed on two supports, the magnet at its center. A goblet is used to keep the hex nuts apart from the magnet and avoid their magnetization. Bottom figure: results. The deformation is linear with the load, and the slope gives the Young modulus.} \label{figure_young}
\end{figure}

Once the magnet/sensor pair is calibrated, the setup is straightforward: a plastic ruler is set horizontally on two supports, the magnet lies on its middle, and the sensor is placed below. To apply a force, objects of known mass are placed on the middle of the ruler. We use metallic hex nuts; to avoid their magnetization (which would interfere with the reading of the distance), a plastic goblet is used to keep them apart from the magnet and still apply their load at the middle of the beam (see top panel of figure~\ref{figure_young}). Nonmagnetic objects would actually be better to apply the load, provided they are heavy enough (such as wood tiles from a construction game for example).

Figure~\ref{figure_young} shows results obtained with a  \SI{50}{\centi\meter}-long,  \SI{4}{\centi\meter}-wide,  \SI{3}{\milli\meter}-thick plastic ruler.  The measured deformation is linear with the load.

With this setup,  small deformations can be described by the linear elasticity theory for a simply supported beam with central load~\cite{refDeformation}, and are given by the equation:
    \begin{equation}
    \delta z = F \frac{L^3}{48 E I}  \quad ,
     \end{equation}
where $\delta z$ is the deformation, $L$ the length of the beam, $F$ is the applied force, $E$ the Young modulus and $I$ is the second moment of area given by:
    \begin{equation}
     I = \frac{b h^3}{12} \quad ,
     \end{equation}
with $b$ the width of the beam, and $h$ its thickness.

The slope of the curve presented in figure~\ref{figure_young} can be used to determine the value of the Young modulus, knowing the dimensions of the ruler and the weight of the load. The weight of 8 hex nuts is \SI{120}{\gram}; the slope gives then a Young modulus of \SI{2}{\giga\pascal}, a reasonable value for plastic~\cite{refYoung}.

Note that the Young modulus can also be obtained by measuring the first harmonic vibration of a cantilever beam, which we proposed to our students as a home assignment project.

\section{Heat Loss}

\textit{Sensor:} thermometer	

\noindent \textit{Experiment:} determining the heat loss of a container

\begin{figure}%[th]
\begin{center}
\includegraphics[width=0.45\textwidth]{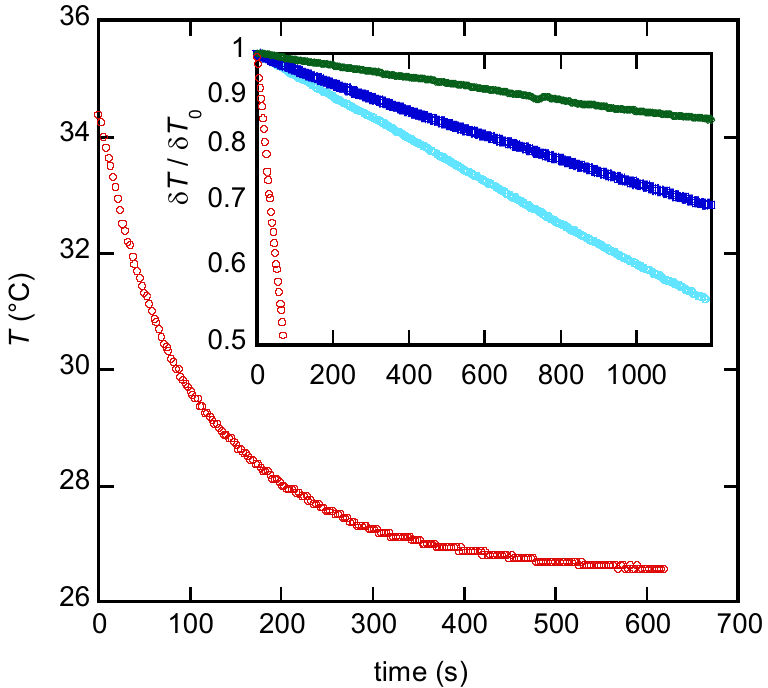}
\end{center}
\caption{Exponential decay of the temperature of \SI{16}{\centi\liter} of water in a plastic goblet initially heated at \SI{34}{\celsius} and put in a  room at \SI{25}{\celsius}. Inset: renormalized temperature $(T-T_{rt})/(T_i-T_{rt})$ in a log scale, as a function of time, for different configurations; from left to right, a goblet immerged in a large volume of water at room temperature (also represented in the main frame of the figure), a goblet in air, a goblet in air with wool cover (hand-knitted sweater) on its sides and below, a goblet fully protected by wool. The corresponding heat loss are respectively \SI{5}{\watt\per\kelvin} ($\tau = \SI{130}{\second}$), \SI{0.34}{\watt\per\kelvin} ($\tau = \SI{2000}{\second}$), \SI{0.2}{\watt\per\kelvin} ($\tau = \SI{3300}{\second}$), and \SI{0.09}{\watt\per\kelvin} ($\tau = \SI{7700}{\second}$).} \label{figure_temperature}
\end{figure}

The aim of this  in-class last session is to recap with the students all they need to know in order to be able to carry on their experimental home assignment. The experiment consists in studying the heat loss of a goblet filled with warm water, and a waterproof Arduino compatible thermometer\cite{refThermometre} is used to follow the temperature of the liquid with time. The main panel of Figure~\ref{figure_temperature} shows the thermal relaxation of water in the goblet from \SI{34}{\celsius} to room temperature when immerged in a large volume of room-temperature water, and also the difference of relaxation time for different goblet isolation. The more the goblet is isolated, the slower the relaxation is.

The physics of thermal relaxation is straightforward: an exponential decay from an initial temperature $T_i$ to room temperature equilibrium $T_{rt}$ can be easily observed:
    \begin{equation}
    T = (T_i - T_{rt})\exp(-t/\tau) + T_{rt} \quad ,
    \end{equation}
where $\tau$ is the characteristic time of the thermal relaxation, due to the balance between thermal inertia and heat loss, $\tau = m c / H$, where $m$ is the mass of water, $c$ the specific heat of water, and $H$ is the heat loss of the system (in \si{\watt\per\kelvin}).

During the in-class session, a single experiment (goblet in air) and considerations on how to determine the heat loss coefficient experimentally fill the session; more elaborate studies comparing the heat loss of different configurations is proposed as a home assignment. Using  renormalized units as shown in the inset of Figure~\ref{figure_temperature}, the heat loss can be determined  through the experimental value of $\tau$. Normalization of $H$ with the surface can be used to compare the results with literature data, such as the isolation guidelines for construction materials.

Various thermodynamic studies can be performed using this sensor, even at home since thermos flasks can be used as good enough calorimeters.

\section{Experimental Homework }

After these training sessions, the students were given an experimental homework, in groups of three or four. Table~\ref{table_projets} shows the list in which our students had to choose a subject. They could borrow an Arduino board and sensors and had to perform the study of their choice at home. They were only given light instructions: no precise protocol but a task and some suggestions on how to tackle it (and the relevant equations  were often given to the students). They had two weeks to present  a written and an oral report on their work. All groups succeeded in producing a satisfying study, albeit with degrees in their quality.

\begin{table*}[ht]
\caption{List of experimental home assignments. The indications ``A'' and ``S'' in the sensor column indicates whether Arduino and/or smartphone sensors can be used.}\label{table_projets}
\centering
\begin{tabular}{p{0.73\linewidth}p{0.25\linewidth}}
Project &  Sensor\\
\hline
		 
Underdamped mechanical oscillations	  & Accelerometer (A) \\
Light transmission through plastic sheets -- Lambert’s Law & Light sensor  (A or S) \\
Supported beam with central load -- Elasticity of materials	 & Hall sensor (A or S)\\
Heat loss of a goblet	 -- Heat transfer & 	Thermometer (A)\\
Rotation of an office chair	 -- momentum conservation	 & Accelerometer (A or S) \\
Resonant modes of a beam -- mechanical resonance	 & Accelerometer (A)\\
Resonant modes of a beam -- mechanical resonance	 & Light sensor (A or S) \\
Attenuation of light with distance	  -- energy conservation	 & Light sensor  (A  or S)\\
Attenuation of light in function of the thickness of a liquid and of the wavelength  -- Lambert's Law	 & Light sensor (A or S) \\
Homemade spectrophotometry to determine strawberry syrup concentration  -- Beer's Law	 & Light sensor (A or S)\\
Measure of the heat capacity of hex nuts (by measuring the decrease of temperature of warm water when the hex nuts are dipped into water)	  & Thermometer (A) \\
Measure of the power of an electric kettle (following the increase of water temperature with time)	  -- specific heat	 & Thermometer (A) \\
Measure of the latent heat of an ice cube (following the temperature of a body of water in which ice cubes are placed)	 & 	Thermometer (A) \\

\hline
\end{tabular}
\end{table*}

A survey was made on the students having followed this class the first year, with 26 answers from 34 participants. Most interestingly, to the multiple-choice question ``Doing an experimental home assignment was:'', the answers were very difficult (0\%), difficult (45\%), easy (50\%), very easy (5\%), and to the multiple-choice question ``this teaching was motivating and I enjoyed learning, I:'' totally disagree (0\%), disagree (0\%), agree (60\%), totally agree (40\%). Even though half of the students didn’t find this exercise easy, they seem to have appreciated doing it. We also asked them the open question ``What do you think this course has brought you?''; the students answers could mainly fall in two categories of roughly equal importance: answers mentioning learning technical skills (on sensors or other experimental technics), and answers mentioning learning about autonomy and scientific methods (on how to construct and lead a study).

From the teacher point of view, this teaching unit was successful and it was decided to renew this teaching the coming year. The interesting aspect of giving an experimental homework compared to an in-class students’ lab, is that students have more time to think for themselves on how to perform a study. This is useful since the main problem that many students faced was a difficulty to understand why to verify a model it is interesting to vary some parameters and measure another quantity that will vary. Many of our students, when given a formula, would try to test it on a given set of parameters, and decide from their single measurement whether the model is valid or not. Discussing with them showed us that they have an idealized vision of science: either it works or it doesn't, and one measurement is enough to determine which. This difficulty can be hidden to a teacher in a classical lab with a protocol for the students to follow: even if they are not convinced of the interest of doing so, the students will perform the expected experiments because the teacher is asking. The training sessions were useful to discuss this aspect with the students prior to the project.

\section{Conclusions}

We used these experiments to teach second-year students how to  study quantitatively a physics phenomenon,  using everyday material and low-cost sensors. These sensors cover a wide range of undergraduate physics: mechanics, optics, thermodynamics, magnetism, and can be used in a variety of other experiments. Smartphone sensors can be used for some of these studies, and present the advantage of ease-of-use;  Arduino- (or any other microcontroller-) compatible sensors   can be more easily tinkered and adapted to more elaborate setups, and offer a wider range of measurements (such as cryogenic temperatures or transport measurements), but can appear more complicated and require some organization to lent the material to the students.

Giving an experimental homework to students (compared to an in-class students’ lab) requires some training sessions beforehand, but gives students a more realistic vision of experimental physics. 
We believe that having students perform experiments in a different context than in a lab is a powerful way of contextualizing physics concepts, but it should be noted that these very same experiments could easily be used for distance learning, lecture experiments or traditional students' labs, at a very low cost. The quality of the results obtained with these low-cost setups is more than sufficient for interesting physics investigations at undergraduate level.

\ack
The authors thank the students of the institut Villebon-\textit{Georges Charpak} who participated to these
projects.  The authors would also like to thank Claire Marrache
for  many fruitful discussions.
This work benefited from a grant
“p\'edagogie innovante” from IDEX Paris-Saclay. The Institut Villebon - \textit{Georges Charpak}  was awarded the Initiative of Excellence in Innovative Training in March 2012 (IDEFI IVICA: 11-IDFI-002), and is supported by the Paris-Saclay Initiative of Excellence (IDEX Paris-Saclay: 11-IDEX-0003).

%\References


\begin{thebibliography}{50}




\bibitem{refArduino} Arduino, \url{https://www.arduino.cc}

\bibitem{refKubinova}  Kub\'inov\'a \v S, and  \v Sl\'egr J 2015, Physics demonstrations with the Arduino board, \PED \textbf{50}, 472.
\bibitem{refGaleriu} Galeriu C,  Edwards S, and  Esper G 2014,  An Arduino investigation of simple harmonic motion, \textit{The Physics Teacher} \textbf{52}, 157.
\bibitem{refAtkin} Atkin K 2016, Construction of a simple low-cost teslameter and its use with Arduino and MakerPlot software,  \PED\textbf{51}, 024001. 
\bibitem{refPetry} Petry C \etal 2016,  Project teaching beyond Physics: Integrating Arduino to the laboratory, in \textit{Technologies Applied to Electronics Teaching (TAEE)},  1--6, IEEE.
\bibitem{refHenaff} Henaff R \etal 2018, A study of kinetic friction: The Timoshenko oscillator, \textit{American Journal of Physics} \textbf{86}, 174.
\bibitem{refBouquet} Bouquet F, Bobroff J, Fuchs-Gallezot M, and Maurines L 2017,  Project-based physics labs using low-cost open-source hardware,  \textit{American Journal of Physics} 85(3), 216. 
\bibitem{refOpenTP} Physics Reimagined (la physique autrement),  \url{http://opentp.fr/en}
\bibitem{refMbed} Arm MBED, \url{https://www.mbed.com}
\bibitem{refAccelerometer} e.g. GoTronic Accelerometres,  \url{https://www.gotronic.fr/cat-accelerometres-1121.htm}
%
\bibitem{refAnalogLightSensor} e.g.  TinkerKit LDR sensors, which use a photoresistor to produce a light-dependent voltage, \url{https://www.vellemanstore.com/en/arduino-tinkerkit-ldr-sensor}
\bibitem{refDigitalLightSensor} e.g.  Adafruit  TSL2561 sensors, \url{https://www.adafruit.com/product/439}
\bibitem{refPhyphox} Staacks S, H\"utz S, Heinke H and Stampfer C 2018, Advanced tools for smartphone-based experiments: phyphox, \PED \textbf{53}, 045009, \url{https://doi.org/10.1088/1361-6552/aac05e}
%Phyphox,  \url{https://phyphox.org}
%
\bibitem{refSpectro1} Anzalone G C ,  Glover A G, and Pearce J M 2013, Open-Source Colorimeter,  \textit{Sensors}, \textbf{13}, 5338. 
\bibitem{refSpectro2}  Nandiyanto A B D, Zaen R, Oktiani R, Abdullah A G, and Riza L S 2018, A Simple, Rapid Analysis, Portable, Low-cost, and Arduino-based Spectrophotometer with White LED as a Light Source for Analyzing Solution Concentration, \textit{Telekomnika} \textbf{16},580.
\bibitem{refDeformation} Hearn E J 1997, \textit{Mechanics of Materials Volume 1: An Introduction to the Mechanics of Elastic and Plastic Deformation of Solids and Structural Materials}, Elsevier.
\bibitem{refYoung} Hearn E J 2013, \textit{Mechanics of Materials Volume 2: An Introduction to the Mechanics of Elastic and Plastic Deformation of Solids and Structural Components}, Elsevier.

\bibitem{refThermometre} e.g.  Grove one-wire temperature sensors, \url{https://www.seeedstudio.com/One-Wire-Temperature-Sensor-p-1235.html}



\end{thebibliography}
\end{document}